\documentclass[article]{JHEP3} 

\usepackage[section] {placeins}
\usepackage{epsfig,multicol}
\usepackage{amsmath}
\usepackage{graphicx}
\usepackage[latin1]{inputenc}
\usepackage{amssymb}
\usepackage{amsmath}
\usepackage{verbatim}
\usepackage{epsfig}
\usepackage{multirow}
\usepackage{booktabs}
\usepackage{cite}
\newcommand{\abs}[1]{\left|#1\right|}

\newcommand\sss{\scriptscriptstyle}
\newcommand{\pt}{p_{\sss T}}
\newcommand{\et}{E_{\sss T}}
\newcommand{\HT}{H_{\sss T}}
\newcommand\clH{{\mathbb H}}
\newcommand\clS{{\mathbb S}}
\newcommand\as{\alpha_{\sss S}}

\newcommand{\HW}{{\sc HERWIG}}

\newcommand{\madloop}{{\sc MadLoop}}

\newcommand{\madfks}{{\sc MadFKS}}

\newcommand{\amcatnlo}{a{\sc MC@NLO}}

\preprint{
 CERN-PH-TH/2011-226 \\
 CP3-11-29 \\
 ZU-TH 16/11
 }
\title{aMC@NLO predictions for $Wjj$ production at the Tevatron}

\author{Rikkert Frederix\\
Institut f\"ur Theoretische Physik, Universit\"at Z\"urich,
Winterthurerstrasse 190,\\ CH-8057 Z\"urich, Switzerland
}
\author{Stefano Frixione%
  \thanks{On leave of absence from INFN, Sezione di Genova, Italy.}\\
  PH Department, TH Unit, CERN, CH-1211 Geneva 23, Switzerland\\
  ITPP, EPFL, CH-1015 Lausanne, Switzerland
}
\author{Valentin Hirschi\\
  ITPP, EPFL, CH-1015 Lausanne, Switzerland
}
\author{Fabio Maltoni\\
  Centre for Cosmology, Particle Physics and Phenomenology (CP3)\\
  Universit\'{e} catholique de Louvain,  B-1348 Louvain-la-Neuve, Belgium
}
\author{Roberto Pittau\\
  Departamento de F\'\i sica Te\'orica y del Cosmos y CAFPE, 
  Universidad de Granada
}
\author{Paolo Torrielli\\
  ITPP, EPFL, CH-1015 Lausanne, Switzerland
}

\abstract{
  We use aMC@NLO to predict the $\ell\nu$ + 2-jet cross section at the
  NLO accuracy in QCD matched to parton shower simulations.
  We find that the perturbative expansion is well behaved for all 
  the observables we study, and in particular for those relevant 
  to the experimental analyses.
  We therefore conclude that NLO corrections to this process cannot be
  responsible for the excess of events in the dijet invariant mass observed
  by the CDF collaboration.
}


\begin{document}


\section{Introduction}
Recently, CDF has reported~\cite{Aaltonen:2011mk} an excess of events in
two-jet production in association with a $W$ boson, in the form of a broad
peak centered at $M_{jj}=144$~GeV in the dijet invariant mass. By now,
i.e.~with a data set corresponding to an integrated luminosity of 7.3
fb$^{-1}$, the excess has reached a statistical significance of 4.1$~\sigma$
w.r.t.~the estimated Standard Model yield. In view of the possible
implications for a BSM physics discovery, this anomaly has attracted a lot
of attention, though it has so far failed to be confirmed by a very similar D0
analysis~\cite{Abazov:2011af}.

One of the major challenges in a measurement of this kind is posed by the need
of reliable predictions and simulations of the processes that contribute to
the observables of interest.  In the CDF and D0 analyses, for instance, such
simulations are typically performed by means of fully exclusive Monte Carlo
programs based on tree-level matrix elements. In the case of multi-jet 
final states in association with weak bosons, a proper merging
procedure~\cite{Catani:2001cc,Krauss:2002up,Alwall:2007fs} between 
multi-parton matrix elements (which give a reliable description of large-angle
and large-energy emissions) and parton shower Monte Carlo's (PSMC's) (which
give a reliable description of small-angle or small-energy emissions) is
employed that allows the generation of inclusive jet samples for all relevant
multiplicities, accurate to the leading order (LO) in perturbative QCD.

Yet, the uncertainties that affect LO predictions can be very large for rates,
and smaller but still discernible for differential distributions. This is the
reason why parton-level NLO and, when possible, NNLO computations of infrared
safe observables are used. Alternatively, and if the statistics is sufficient,
control data samples are employed.  For example, a theoretical analysis based
on the NLO computation of the SM yield for $\ell + 2$ jets + missing
transverse energy (which with the cuts used by CDF and D0 gets contributions
from, in order of importance, $Wjj$, $Zjj$,
$WW$, $t\bar t$, single-$t$ and $WZ$ production) has recently
appeared~\cite{Campbell:2011gp}. It has been shown that indeed the $Wjj$
process gives by far the dominant contribution, and that the NLO QCD 
corrections are small. Unfortunately, even though more accurate from the
theoretical point of view, such small-multiplicity, parton-level calculations
cannot be directly compared to experimental analyses, since this would
require events with high-multiplicity, fully-fledged hadronic final states.

In order to obtain predictions that are both accurate and employable in
experimental analyses, an NLO calculation needs to be consistently matched
to a PSMC. This can be currently achieved with the
MC@NLO~\cite{Frixione:2002ik} or POWHEG
methods~\cite{Nason:2004rx,Frixione:2007vw}. It is interesting to note that
out of the processes listed above for the signature $\ell + 2$ jets +
missing transverse energy, only the $Wjj$ and $Zjj$ contributions
are not available in either of these frameworks. Given that the
cross section of the latter process (within the experimental cuts
adopted by CDF and D0) is smaller than that of the former by more
than one order of magnitude, it is more urgent and highly desirable 
to have the best possible theoretical predictions for $Wjj$ 
production, which is a fairly challenging task. The complexity 
stems not only from the NLO computation itself, but also from its
subsequent matching with parton showers, where the technical difficulties
arise mainly from the presence of phase-space singularities at the Born level,
which need to be cut-off.  While this problem has already been faced in the
POWHEG implementation of dijet~\cite{Alioli:2010xa} and 
$Wj/Zj$~\cite{Alioli:2010qp} production, it is significantly simpler in these
cases: a $\pt$ cut on the ``recoil'' system (one parton in dijet, and the
vector boson in $Wj/Zj$ production) is sufficient to get rid of the
divergences of the Born matrix elements. On the other hand, $Wjj$ production
features a final-state three-body (of which two light partons) kinematic
configuration already at the Born level, which renders the cutting-off of the
singularities highly non trivial.  
In fact, the kinematics of $Wjj$ production is sufficiently involved
to provide a proof that, if a successful matching of the NLO results with
parton showers can be achieved, the same kind of matching technique can be
applied to larger final-state multiplicities, without encountering any new
problems of principle.

In this paper, we compute the NLO QCD corrections to the process
$p\bar{p}\to\ell \nu jj$\footnote{The mass of the charged lepton 
$\ell$ is set equal to zero. Furthermore, since we do not compare our
predictions to data here, it is sufficient to consider only positively-charged 
leptons of one flavor.} and, for the first time and in a fully-automated way, consistently 
match them to the \HW\ parton shower~\cite{Marchesini:1991ch,Corcella:2000bw,
Corcella:2002jc}
according to the MC@NLO formalism~\cite{Frixione:2002ik}, as implemented 
in the \amcatnlo\ program~\cite{amcatnlo}. One-loop corrections are obtained 
with \madloop~\cite{Hirschi:2011pa}, which is based on the OPP reduction
method~\cite{Ossola:2006us} and on its implementation in 
CutTools~\cite{Ossola:2007ax}. All the other contributions 
to the parton-level NLO cross section are dealt with
by \madfks~\cite{Frederix:2009yq}, which is based on the FKS subtraction
method~\cite{Frixione:1995ms}, and takes care of determining the MC
counterterms needed in the MC@NLO approach.
Throughout the paper, we often refer to ``the $W$ boson'' or to 
``$Wjj$ production''; this is only for the sake of brevity,
since we actually deal with the leptonic process mentioned before,
and thus doing we fully retain the information on production and decay spin
correlations and off-shell effects.

We begin by showing that the cutting-off of Born-level singularities
(which is an arbitrary procedure) has no impact
on the predictions in the kinematic regions of interest.
We also show that NLO corrections are moderate, and
depend mildly on the kinematics. We conclude by presenting our predictions
for the dijet invariant mass, closely following the CDF analysis.

\section{Method and validation}
The $Wjj$ NLO cross section receives contributions from processes with
$W\!+\!2$-parton and $W\!+\!3$-parton final states; although these diverge
when independently integrated over the phase space, their combination into any
infrared-safe observable is finite, thanks to the KLN and factorization
theorems. For this to happen, it is a crucial condition that there be two
observable jets in the final state. Although such a condition can be easily
included in the definition of the short-distance cross sections, this is not
the way one follows nowadays. A much-preferred option (and the only one which
is viable when matching to PSMC's) is that of imposing jet cuts at the very
last step of the computation (the physics analysis), since this gives one the
flexibility of e.g.~using several jet-finding algorithms in parallel. It
should be clear, however, that some cuts (called generation cuts henceforth)
must still be imposed at the level of short-distance cross sections, which
otherwise would diverge upon integration, as mentioned before. Generation cuts
are therefore a technical trick that allow one to work with finite quantities;
the idea is that kinematic configurations that do not pass these cuts would
anyhow not contribute to the observable cross sections, which is what permits
one to discard them; in other words, cross sections are not biased by
generation cuts.

Unfortunately, it is not straightforward to prove that indeed
physical observables are unbiased, which constitutes a necessary and very
strong consistency check of one's computation. An analytic proof not being
viable, one exploits the fact that generation cuts are arbitrary. Hence, one
imposes several generation cuts, and then verifies that in the kinematic
regions of interest physical observables do not depend on them. 
This opens the question of how to define generation cuts, and
it is obvious that a necessary condition is that they must be looser 
than the loosest of the set of cuts imposed in the physics analysis. 
When performing a perturbative calculation at the parton level, 
it is quite easy to understand whether generation cuts are sufficiently
loose. This is because generation and analysis cuts are imposed on kinematic
configurations that have the {\em same} multiplicities and particle contents.
Things are significantly more complicated when one matches
matrix-element computations with parton showers; the latter will in fact
generally increase the final-state multiplicities w.r.t.~those relevant
to short-distance cross sections, and the relationship between the
quantities being cut at the generation and analysis level becomes blurred.
The upshot of this is the following: when considering the matching
with parton showers, generation cuts are typically softer than those
one would need if only performing perturbative parton-level computations,
and they affect larger kinematic ranges than in the latter case.

In order to address this (among others) problem, at the LO one ``merges''
different parton multiplicities in a way consistent with parton
showers~\cite{Catani:2001cc,Krauss:2002up,Alwall:2007fs}. Although a 
generalization of these procedures to NLO is in its 
infancy~\cite{Nagy:2005aa,Giele:2007di,Lavesson:2008ah,Giele:2011cb,
Hamilton:2010wh,Hoche:2010kg,Alioli:2011nr}, we may observe that when 
the merging at the LO is restricted to processes whose multiplicities
differ by one unit, then one is actually dealing with (a subset of) the 
matrix elements used in the well-established NLO-PSMC matching
procedures such as MC@NLO. Hence, one may anticipate that unphysical
effects, the reduction of whose impact necessitates a merging procedure
at the LO, are smaller in the context of matched NLO computations
of a given multiplicity. We shall later see an explicit example
of this fact.

To conclude this discussion, we mention that, although there is ample freedom
in the choice of generation cuts, in practice it is convenient to employ the
same jet-finding algorithm at the matrix element level as in the physics
analysis, since this renders it a bit easier the task of applying generation
cuts which are looser than the analysis ones.

As a technical aside, we point out that the MC@NLO formalism does not 
require modifications in order to be applied to processes whose Born 
contribution is divergent, and one simply imposes generation cuts 
when computing MC@NLO short-distance cross sections, fully analogously
to what is done at the LO. Using the results of 
ref.~\cite{Frixione:2002ik}, it is easy to show~\cite{amcatnlo}
that this should be done in the following way. All contributions to
$\clS$ events, and the MC counterterms relevant to $\clH$ events,
are cut according to the corresponding Born configuration, while the
real-emission contributions to $\clH$ events are cut according to 
the corresponding fully-resolved configuration.

Our predictions are obtained with the electroweak parameters
reported in table~\ref{tab:params}. For the (N)LO computations we
use the MSTW(n)lo200868cl~\cite{Martin:2009iq}~PDFs, which 
also set the value of $\alpha_{\sss S}(M_{\sss Z})$. The
renormalization and factorization scales are chosen equal to $\HT/2$, 
with \mbox{$\HT=\sum_i  p_{{\sss T},i}+\sqrt{\pt^2(\ell\nu)+M^2(\ell\nu)}$}. 
The sum here runs over all final-state QCD partons, and all the quantities 
that appear in the definition of $\HT$ are computed at the matrix-element 
level, i.e., before showering. We have not included the simulation of the underlying event in our predictions.

\begin{table}
\begin{center}
\begin{tabular}{ll|ll}
Parameter$~~~~$ & value & Parameter$~~~~$ & value
\\\hline
$m_{W}$ & 80.419 & $\Gamma_W$    & 2.0476 
\\
$G_F$   & $1.16639\!\cdot\!10^{-5}$ &$\alpha^{-1}$ & 132.50698 
\\
$m_t$   & 174.3 &  $m_{Z}$ & 91.118 
\\
$\as^{({\rm NLO})}(m_{Z})$  & 0.12018 & 
$\as^{({\rm LO})}(m_{Z})$   & 0.13939 
\end{tabular}
\end{center}
\caption{\label{tab:params}
  Settings of physical parameters used in this work,
  with dimensionful quantities given in GeV.
}
\end{table}

We define jets by means of the anti-$k_T$ algorithm~\cite{Cacciari:2008gp} 
with $R=0.4$, as implemented in FastJet~\cite{Cacciari:2005hq}. 
Generation cuts are imposed by demanding the presence of at least two
jets at the hard-subprocess level (hence, at this stage the inputs to the 
jet-finding algorithm are two- or three-parton configurations).
All jets thus found are required to have either $\pt\!>\!5$~GeV
or $\pt\!>\!10$~GeV. The short-distance cross sections defined
with these cuts are used to obtain unweighted events as customary
in MC@NLO. Such events are then showered by \HW, and the resulting
hadronic final states are used to reconstruct about sixty
observables (involving leptons, jets, lepton-jet, and jet-jet correlations)
for each of the two generation $\pt$ cuts mentioned above. 
These observables are organized in three classes, each being associated 
with jets\footnote{We stress that such jets are now reconstructed 
by clustering all stable final-state hadrons that emerge from the shower.}
defined by imposing their transverse momenta to be larger than 10, 25, 
and 50~GeV; these conditions will be called analysis cuts henceforth.
We finally check that the tighter the analysis cuts, the smaller the
difference between the results obtained with the two generation cuts.

\begin{figure}[t!]
  \begin{center}
        \epsfig{file=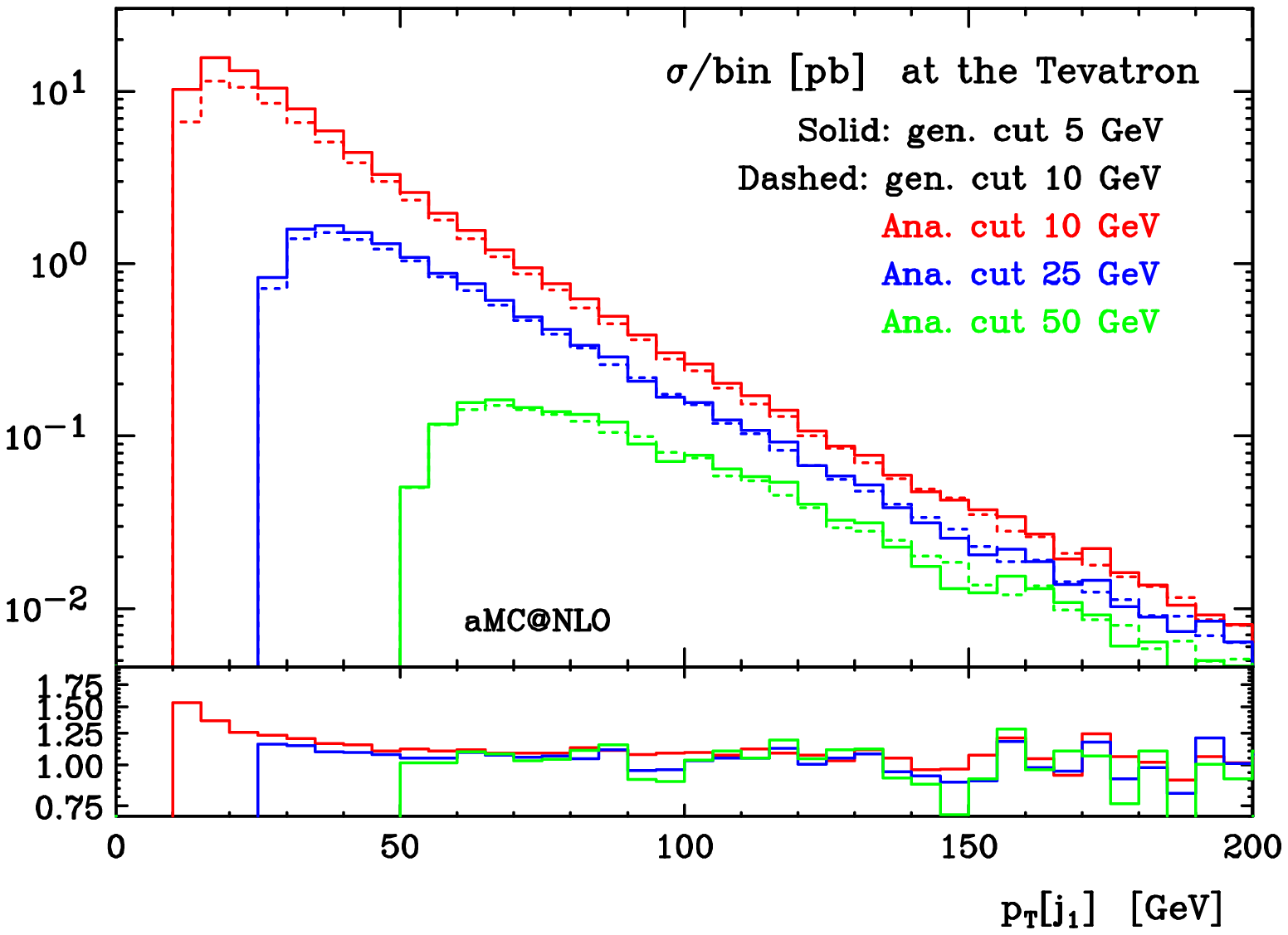, width=0.60\textwidth}
  \vspace{3pt}
        \epsfig{file=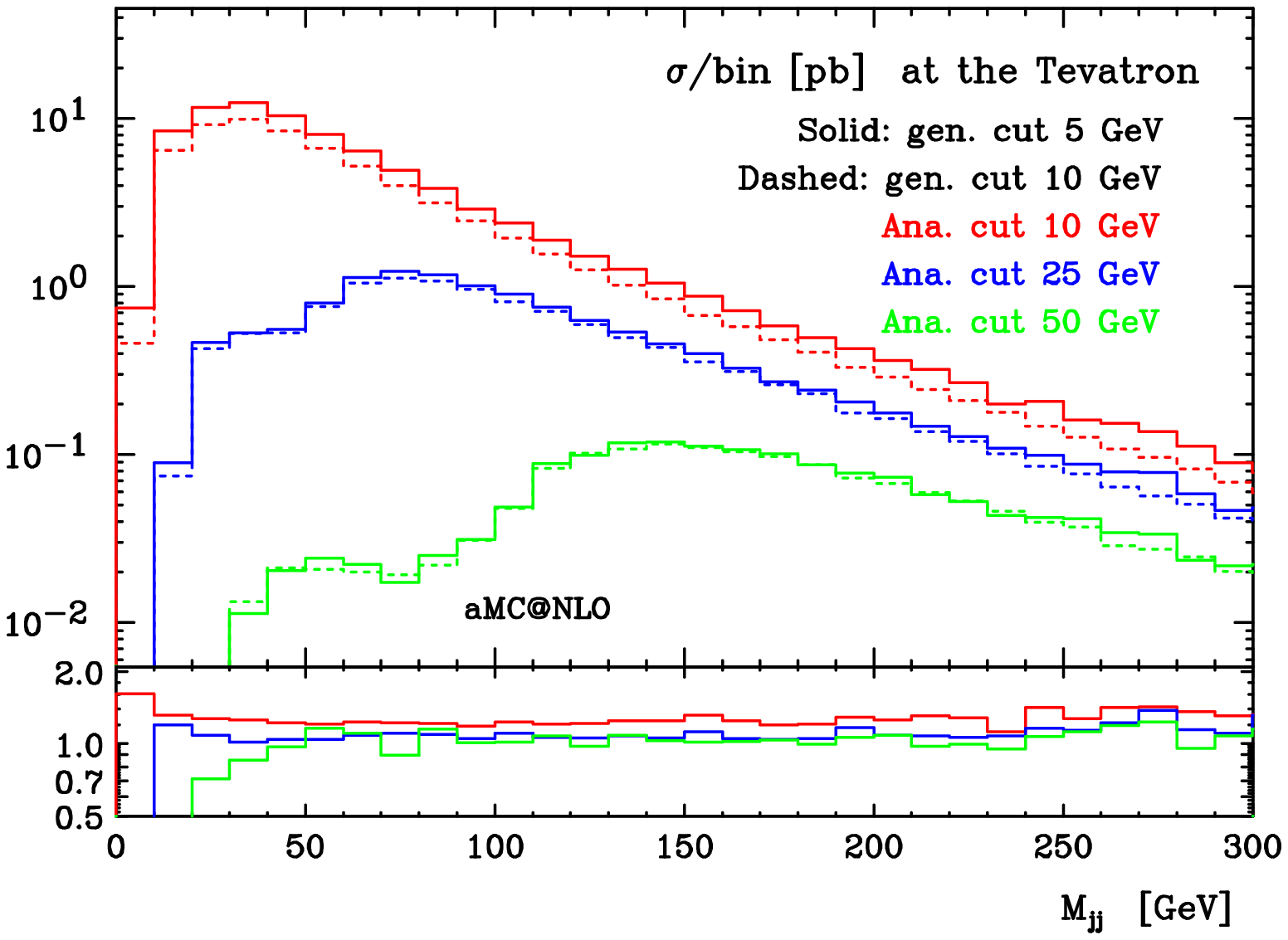, width=0.60\textwidth}
  \vspace{3pt}
        \epsfig{file=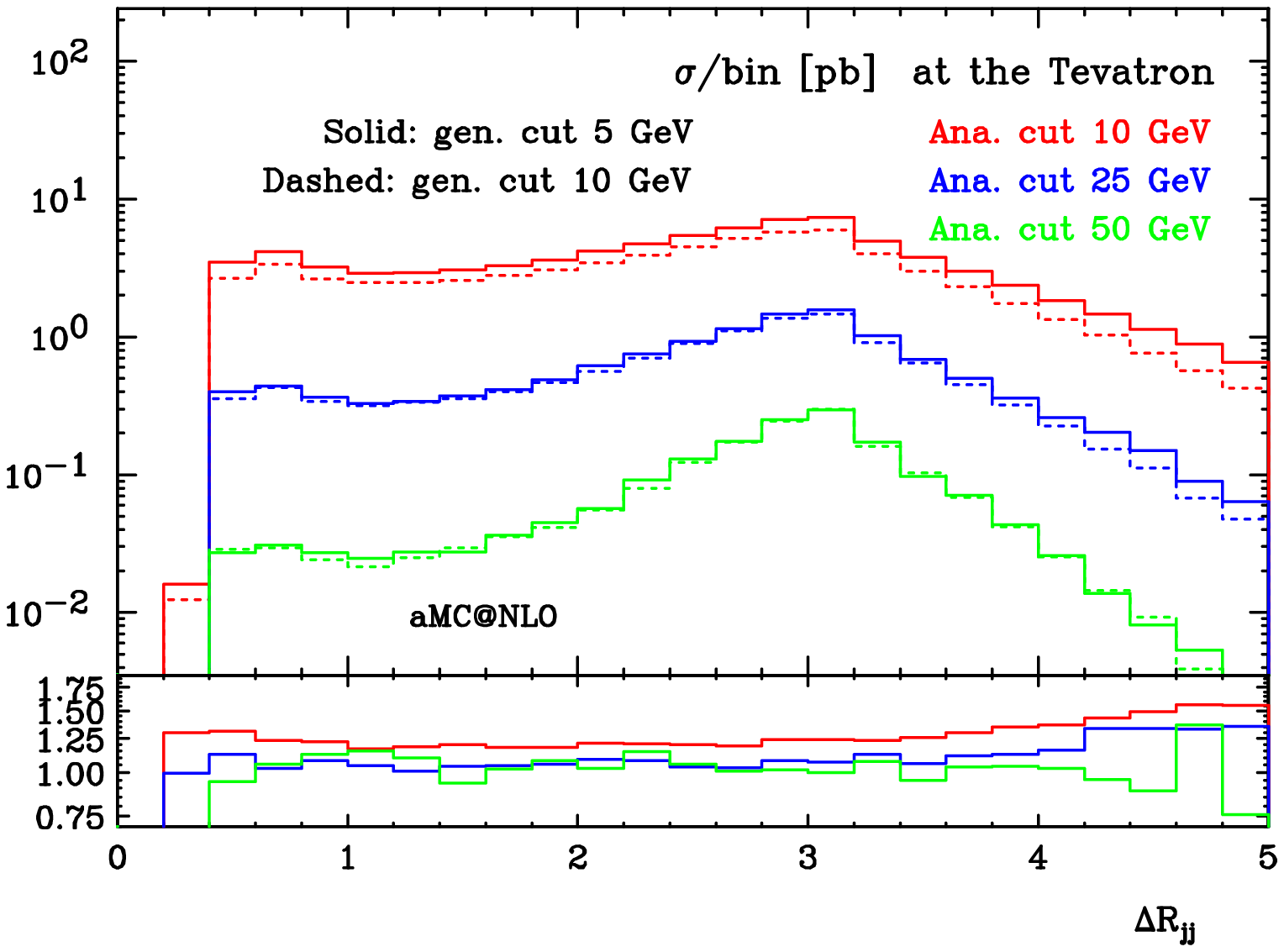, width=0.60\textwidth}
  \end{center}
  \vspace{-18pt}
  \caption{Transverse momentum of the hardest jet (upper plot), 
   invariant mass of the pair of the two hardest jets (middle plot)
   and distance between the two hardest jets in the $\eta-\varphi$ plane
   (lower plot), in $Wjj$ events and as predicted by \amcatnlo. See the text 
   for details.
}
\label{fig:all}
\end{figure}

As an example of the outcome of this exercise, we present in
fig.~\ref{fig:all} the transverse momentum of the hardest jet, 
the dijet invariant mass, and the $\Delta R$ separation
between the two hardest jets. In the main frame of each plot
there are six histograms: the three solid ones correspond to generation cuts
$\pt=5$~GeV, while the three dashed ones correspond to generation cuts
$\pt=10$~GeV.  The upper (red), middle (blue), and lower (green) pairs of
histograms are obtained with the analysis cuts $\pt\!=\!10$, $25$, and $50$~GeV
respectively. The lower insets display three curves, obtained by taking
the ratios of the $\pt\!=\!5$~GeV generation-cut results over the 
$\pt\!=\!10$~GeV generation-cut results, for the three given analysis cuts
(in other words, these are the ratios of the solid over the dashed
histograms). Fully-unbiased predictions are therefore equivalent to
these ratios being equal to one in the kinematic regions of interest.

Inspection of fig.~\ref{fig:all}, and of its analogues
not shown here, allows us to conclude that the results follow the expected
pattern: when one tightens the analysis cuts, the bias due to the
generation cuts is reduced, and eventually disappears. Although all
observables display this behaviour, the precise dependence on generation
cuts is observable-specific; the three cases of 
fig.~\ref{fig:all} have been chosen since they are representative 
of different situations. The transverse momentum of the hardest 
jet shown in the upper plot of fig.~\ref{fig:all} 
is (one of) the very observable(s) on which generation cuts are imposed.
Therefore, as one moves towards large $\pt$'s, one expects the bias
due to generation cuts to decrease, regardless of values of the $\pt$
cut used at the analysis level. This is in fact what we see. Still,
a residual dependence on generation cuts can be observed at relatively
large $\pt$'s for looser analysis cuts; this could in fact be anticipated,
since the events used here are $Wjj$ ones -- hence, the next-to-hardest
jet will tend to have a transverse momentum as close as possible to
the analysis $\pt$ cut, and thus to the region affected by the generation
bias in the case of looser analysis cuts. The dijet invariant mass, 
shown in the middle  plot of fig.~\ref{fig:all}, tells a slightly different
story. Namely, the hard scale associated with this observable is not
in one-to-one correspondence with that used for imposing the analysis
cuts, at variance with the $\pt$ of the hardest jet discussed previously.
Hence, the effects of the generation-level cuts are more evenly distributed
across the whole kinematical range considered, as can be best seen from
the lower inset. Essentially, the bias here amounts largely to a normalization
mismatch, which disappears when tightening the analysis cuts.
Finally, the $\Delta R$ distribution, presented in the lower part of 
fig.~\ref{fig:all}, is representative of a case where both shapes and
normalization are biased. There is a trend towards larger biases at large
$\Delta R$, which is understandable since this region receives the most
significant contributions from large-rapidity regions, where the transverse
momenta tend to be relatively small and hence closer to the bias region.

We conclude this section with some further comments on validation
exercises. Firstly, we started by testing the whole machinery 
in the simpler case of $Wj$ production. Although, as was discussed
before, for this process generation cuts may be imposed on $\pt(W)$,
we have chosen to require the presence of at least one jet with
a transverse momentum larger than a given value, so as to mimic
the strategy followed in the $Wjj$ case. Secondly, we have checked 
that we obtain unbiased results by suitably changing the jet-cone 
size. Thirdly, we have exploited the fact
that the starting scale of the shower is to some extent arbitrary,
and the dependence upon its value is very much reduced in the
context of an NLO-PSMC matched computation. As was discussed in
ref.~\cite{Torrielli:2010aw}, in MC@NLO the information on the
starting scale is included in the MC counterterms, and the 
independence of the physical results of its value constitutes
a powerful check of a correct implementation. We have verified
that this is indeed the case.

\section{$Wjj$ production at the Tevatron}
The hard events obtained with the generation cuts described above
can be used to impose the selection cuts employed by the CDF 
collaboration~\cite{Aaltonen:2011mk}. The latter are as follows
(where with ``lepton'' we always mean the charged one):
\begin{itemize}
\item minimal transverse energy for the lepton: $\et(\ell) > 20$~GeV;
\item maximal pseudorapidity for the lepton: $|\eta(\ell)| < 1$;
\item minimal missing transverse energy: $E\!\!\!\!/_{\sss T} > 25$~GeV;
\item minimal transverse $W$-boson mass: $M_{\sss T}(\ell\nu) > 30$~GeV;
\item jet definition: JetClu algorithm with 0.75 overlap and $R=0.4$;
\item minimal transverse jet energy: $\et(j) > 30$~GeV;
\item maximal jet pseudorapidity: $|\eta(j)| < 2.4$;
\item minimal jet pair transverse momentum: $\pt(j_1j_2) > 40$~GeV;
\item minimal jet-lepton separation: $\Delta R(\ell j)>0.52$;
\item minimal jet-missing transverse energy separation:
 $\Delta\phi(E\!\!\!\!/_{\sss T}j)>0.4$;
\item hardest jets close in pseudorapidity: $\abs{\Delta\eta(j_1j_2)}<2.5$;
\item lepton isolation: transverse hadronic energy smaller than 10\% of 
  the lepton transverse energy in a cone of $R=0.4$ around the lepton.
\item jet veto: no third jet with $\et(j)>30$~GeV and $\abs{\eta(j)}<2.4$;
\end{itemize}
These cuts (and their analogues in the D0 analysis~\cite{Abazov:2011af},
which give very similar results in the ``signal'' region)
are tighter than the $\pt=25$~GeV analysis cut previously discussed.
Since the latter was seen to give unbiased results in the central
rapidity regions relevant here, we deem our approach safe. 
The cuts reported above (which we dub ``exclusive'') have also been
slightly relaxed by CDF (see~\cite{CDFweb}), by accepting events
with three jets or more in the central and hard region -- this amounts
to not applying the jet-veto condition reported in the last bullet above;
we call these cuts ``inclusive''. 

\begin{figure}[t!]
  \begin{center}
        \epsfig{file=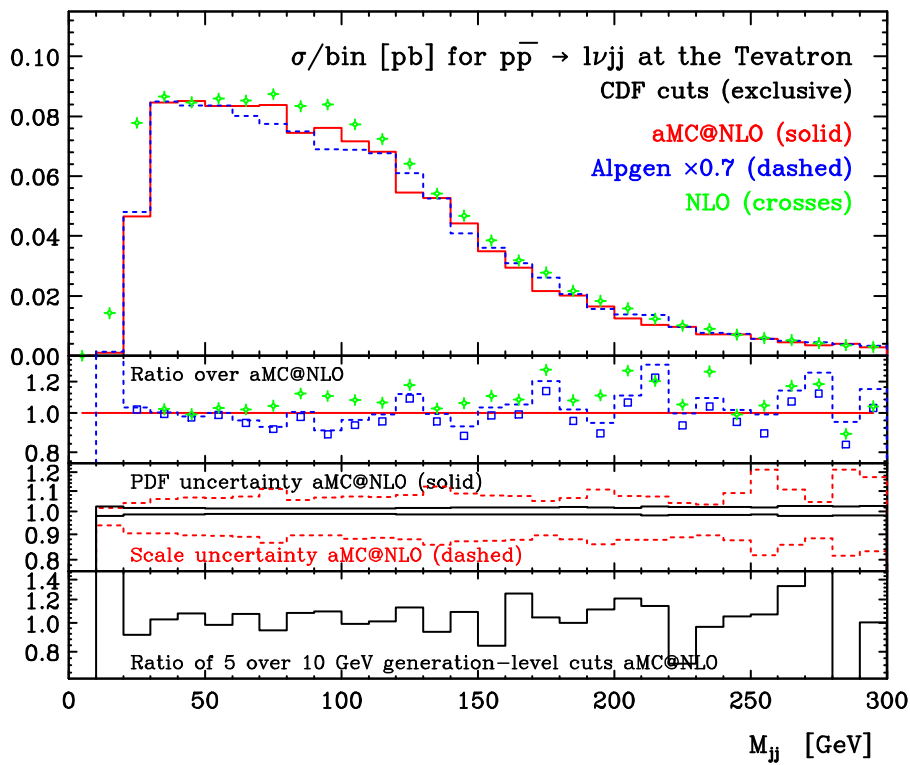, width=0.79\textwidth}
  \end{center}
  \vspace{-20pt}
  \caption{Invariant mass of the pair of the two hardest jets, with
  CDF/D0 exclusive cuts. See the text for details.}
  \label{fig:CDFD01}
\end{figure}

\begin{figure}[t!]
  \begin{center}
        \epsfig{file=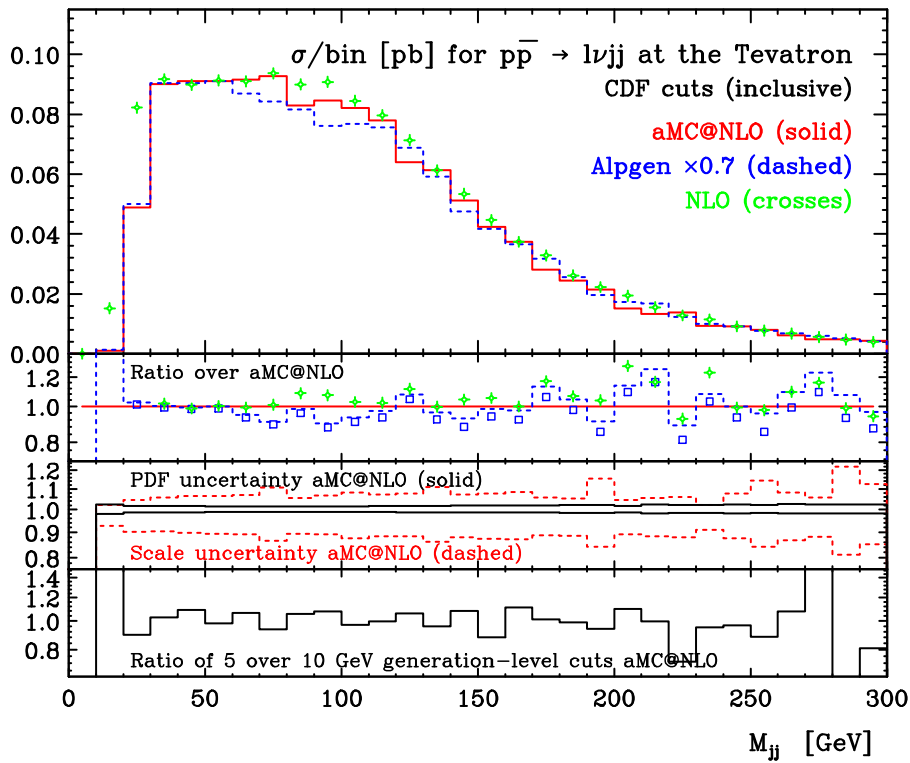, width=0.79\textwidth}
  \end{center}
  \vspace{-20pt}
  \caption{Invariant mass of the pair of the two hardest jets, with
  CDF/D0  inclusive cuts. See the text for details.}
  \label{fig:CDFD02}
\end{figure}

In addition to the \amcatnlo\ predictions, we have performed parton-level
LO and NLO computations. Finally, we have showered events obtained by 
unweighting LO matrix elements as well. As is well known, the latter
case is potentially plagued by severe double-counting effects which,
although formally affecting perturbative coefficients of order higher
than leading, can be numerically dominant. We have indeed found that
this is the case for the cuts considered here: predictions obtained
with generation cuts $\pt\!=\!5$~GeV and $\pt\!=\!10$~GeV differ by
30\% or larger for total rates (shapes are in general better agreement),
even for the analysis cut of $\pt\!=\!50$~GeV. We have therefore opted
for using a matched LO sample, which we have obtained with
Alpgen~\cite{Mangano:2002ea} interfaced to \HW\ through the MLM 
prescription~\cite{Alwall:2007fs}. In order to do this, we have
generated $W+n$~parton events, with $n=1,2,3$. The dominant contribution
to $Wjj$ observables is due to the $n=2$ sample, but that of $n=3$ is not
negligible. The size of the $n=1$ contribution is always small, and
rapidly decreasing with dijet invariant masses; it is thus fully safe
not to consider $W+0$~parton events.

In figs.~\ref{fig:CDFD01} and \ref{fig:CDFD02} we present our predictions for the invariant
mass of the pair of the two hardest jets with exclusive and inclusive cuts, respectively.
The three histograms in the main frames are the aMC@NLO (solid red), 
Alpgen+MLM (dashed blue), and NLO parton level (green symbols) predictions. 
The two NLO-based results are obtained with the $\pt\!=\!10$~GeV generation 
cuts. The Alpgen+MLM curves have been rescaled to be as close as possible
to the NLO ones, since their role is that of providing a prediction for
the shapes, but not for the rates (incidentally, this is also what is
done in the experimental analyses when control samples are not available).
The upper insets show the ratios of the Alpgen+MLM and NLO results over
the aMC@NLO ones. The middle insets display the fractional scale (dashed red) and PDF
(solid black) uncertainties given by \amcatnlo, computed with the
reweighting technique described in ref.~\cite{Frederix:2011ss}.
The lower insets show the ratios of the aMC@NLO
results obtained with the two generation cuts, and imply that indeed there
is no bias due to generation cuts. We have also checked that removing the lepton isolation cut
does not change the pattern of the plots, all results moving consistently
upwards by a very small amount.

By inspection of figs.~\ref{fig:CDFD01} and \ref{fig:CDFD02}, we can conclude that the three
predictions agree rather well, and are actually strictly equivalent,
when the theoretical uncertainties affecting \amcatnlo\ are taken into 
account (i.e., it is not even necessary to consider those relevant
to Alpgen+MLM and parton-level NLO).
This is quite remarkable, also in view of the fact that the
dominant contribution to the latter, the scale dependence, amounts to
a mere $(+10\%,-15\%)$ effect. We have verified that such a dependence
is in agreement with that predicted by MCFM~\cite{Campbell:2002tg}.

In spite of their being not significant for the comparison with data, 
it is perhaps interesting to speculate on the tiny differences between
the central \amcatnlo, Alpgen+MLM, and NLO predictions.
The total rates given by \amcatnlo\ and NLO are close but not
identical; this is normal, and is a consequence of the fact that
the kinematical distributions in the two computations are different,
and thus differently affected by the hard cuts considered here.
More interestingly, the $M_{jj}$ distribution predicted at the NLO
is (very) slightly harder than that of \amcatnlo, especially in
the case of exclusive cuts. This is best seen in the upper insets
of figs.~\ref{fig:CDFD01} and \ref{fig:CDFD02}, and is due to the fact that the fraction of 
events with a third central and hard jet is larger in \amcatnlo\ than 
at the parton-level NLO. This argument applies also to the case of
inclusive cuts. In fact, by requiring the two hardest
jets to have a large invariant pair mass, and given the presence of
a $W$ boson, one forces extra QCD radiation to be fairly soft,
since relatively-hard radiation is strongly suppressed by the damping
of the PDFs at large Bjorken $x$'s. This effectively imposes a veto-like
condition on the events, which however, at $M_{jj}\simeq 300$~GeV, is still
larger than the explicit 30~GeV one imposed by CDF; hence, NLO predictions
for inclusive cuts are slightly harder than the \amcatnlo\ ones, but
less than in the case of exclusive cuts. We point out that a veto
on the third jet (be it explicit or effective) introduces a new mass
scale in the problem, whose ratio over $M_{jj}$ may grow large.
In such a situation, the resummation of large logarithms performed 
by the shower constitutes an improvement over fixed-order results.
Given the level of agreement we find here, we can conclude the resummation
effects are still fairly marginal.

As far as the comparison between the central \amcatnlo\ and Alpgen+MLM 
predictions is concerned, this is affected by the choice of the 
hard scales, which are different in the two codes: in Alpgen,
the transverse $W$-boson mass is adopted (the renormalization scale is
then effectively redefined through the reweighting of the matrix
elements by $\as$ factors, which is specific of the merging
procedure~\cite{Catani:2001cc}). In spite of this, the agreement
between the two results is quite good, with Alpgen+MLM being slightly
harder than \amcatnlo\ (this effect being of the same order or smaller
than that observed with parton-level NLO results). We have also compared
Alpgen+MLM with \amcatnlo, by setting the hard scales in the latter
equal to the transverse $W$-boson mass\footnote{Note that, since we determine
the scale dependence through the reweighting technique of 
ref.~\cite{Frederix:2011ss}, we do not need to run \amcatnlo\ a second time.}. 
The ratio of these two results is shown  as open boxes in 
the upper inset of figs.~\ref{fig:CDFD01} and \ref{fig:CDFD02}, whence one sees a marginal
improvement in the agreement between the two predictions w.r.t.~the
case corresponding to $\mu=\HT/2$, which is our \amcatnlo\ default.
We finally stress again that the MLM prescription is crucial to get
rid of double-counting effects in LO samples. While double counting
is guaranteed not to occur at the NLO in MC@NLO, it can still affect 
terms of ${\cal O}(\as^4)$ and beyond. Although we did not see any 
evidence of these in the form of generation-cut dependence, we
have also heuristically extended the MLM prescription to NLO, by 
requiring the two hardest jets after shower to be matched with two
jets reconstructed at the hard-subprocess level (where they play the
same roles as the partons in the original MLM matching). This prescription
has had no visible effect on our results. Although this is a process-dependent
conclusion, it confirms the naive expectation that NLO-PSMC matching is
less prone to theoretical systematics than its LO counterpart, and
suggests that a reduction of the dependence upon unphysical merging
parameters can be achieved by extending the CKKW or MLM procedures
to the NLO.

\section{Conclusions}
In this paper, we have presented the automated computation of the
$Wjj$ cross section to the NLO accuracy in QCD, and its matching
to parton showers according to the MC@NLO formalism. This is the
first time that a process of this complexity has been matched to
an event generator beyond the LO. We believe this is
significant not only as a phenomenological result, but also in
view of the fact that it is also the first time that the MC@NLO
prescription has been applied to a process that requires the 
presence of cutoffs at the Born level in order to prevent phase-space
divergences from appearing. In fact, the structure of such divergences 
in $Wjj$ production is sufficiently involved to provide evidence that
no new problems of principle are expected in the application of
MC@NLO to processes with even larger final-state multiplicities.

We have given predictions for the dijet invariant mass in $Wjj$ events, using
the same cuts as CDF and D0 in the signal region. Perturbative, parton-level
results agree well with those obtained after shower, and we do not observe any
significant effects in the shape of distributions due to NLO corrections,
which therefore cannot be responsible for the excess of events observed by the
CDF collaboration.

\section{Acknowledgments}

We would like to thank Johan Alwall, Michelangelo Mangano and Bryan
Webber for useful discussions. S.~F. is indebted to Michelangelo Mangano
for his assistance in running Alpgen. This research has been supported by
the Swiss National Science Foundation (SNF) under contract
200020-138206, by the Belgian IAP Program, BELSPO P6/11-P and the IISN
convention 4.4511.10, by the Spanish Ministry of education under
contract PR2010-0285. F.M.~and R.P.~thank the
financial support of the MEC project FPA2008-02984 (FALCON).
R.F.  and R.P. would like to thank the KITP at UCSB for the kind hospitality offered 
while an important part of this work was being accomplished.

\bibliography{w2j}

\end{document}